\documentclass[prd,amsmath,amssymb,english,floatfix,nofootinbib]{revtex4}
\usepackage{graphicx}
\usepackage{dcolumn}
\usepackage{bm}
\usepackage{epsfig}
\usepackage{babel}
\usepackage{color}
\usepackage{xspace}
\usepackage{subfigure}

\newcommand{\pth}{\hat{p}_{\perp}^{\;min}}

\newcommand{\CHECK}[1]{\textbf{\color{red}[#1]}\xspace}
\newcommand{\chk}[1]{\CHECK{CHECK THIS!}}

\begin{document}

\thispagestyle{empty}

~\\[20mm]
\centerline{\Large\bf Double parton interactions as a background to associated {\it HW}}\\[1mm]
\centerline{\Large\bf production at the Tevatron}
~\\[5mm]

\centerline{Dmitry Bandurin$^{1}$, Georgy Golovanov$^{2}$, Nikolai Skachkov$^{2}$} 
~\\[0mm]
\centerline{$^1$ {\it Department of Physics, Florida State University, Tallahassee, FL 32306}}
\centerline{$^2$ {\it Joint Institute for Nuclear Research, Dubna, Russia, Joliot-Curie 6, 141980}}
~\\[5mm]
\centerline{\bf Abstract}
In this paper we study events with $W$+jets final state, produced
in double parton (DP) interactions, as a background to the associated Higgs boson ($H$)
and $W$ production, with $H\to b\bar{b}$ decay, at the Tevatron. 
We have found that the event yield from the DP background can be quite sizable,
which necessitates a choice of selection criteria to separate the $HW$ and DP production processes.
We suggest a set of variables sensitive to the kinematics of DP and $HW$ events.
We show that these variables, being used as an input to the artificial neural network, 
allow one to significantly improve a sensitivity to the Higgs boson production.

\newpage

\clearpage

\section{Introduction}\markright{\thesubsection\hspace{1em}Introduction}{} 
\label{Sec:intro}

  A significant amount of experimental data, ranging from ISR energies \cite{AFS} through the
  SPS \cite{UA2} to the Tevatron \cite{CDF93, CDF97, E735, D003,D0_2010,D0_2011}, and even
  to photoproduction at HERA \cite{ZEUS,H1}, shows clear evidence of hard
  jets produced from multiple parton interactions (MPI).
  Specifically, in the Tevatron Run I and Run II studies, 4-jet \cite{CDF93} and $\gamma+3$-jet
  events  \cite{CDF97,D0_2010} have been considered with jet $p_T\gtrsim 5-15$ GeV,
  and the fraction of events occurring due to 
  double parton (DP) interactions have been measured. Those fractions varied 
  depending on the final state and the jet transverse momentum ($p_T$)
  of the second parton interaction.
  The fraction measured using 4-jet final state is 
  found to be $5.5\%$ for jet $p_T>25$ GeV \cite{CDF93}.
  The fractions obtained from the $\gamma+3$-jet production range from
  $51.3\%$ for the second (ordered in $pT$) and third jet $p_T$ in the
  interval $5-7$ GeV\footnote{In this measurement jet $p_T$ is raw, 
  i.e. uncorrected for the energy losses \cite{CDF97}.} \cite{CDF97} to $47\%-22\%$ for the second jet
  $p_T$ within $15-30$ GeV \cite{D0_2010}.


  Those experiments have also measured the effective cross section $\sigma_{\rm eff}$,
  an important 
  parameter that contains information about the parton spatial density inside the (anti)proton:
  $\sigma_{\rm eff} = 12.1_{-5.4}^{+10.7}$ mb in the 4-jet production in CDF \cite{CDF93}, 
  $\sigma_{\rm eff} = 14.5$$\pm$$1.7_{-2.3}^{+1.7}$ mb 
  and $\sigma_{\rm eff}=16.4\pm0.3\pm2.3$ mb
  in the $\gamma+3$-jet productions in CDF \cite{CDF97} and D0 \cite{D0_2010}, respectively.
  This parameter allows the calculation of a DP cross section  $\sigma_{DP}$
  for any pair of partonic processes $A$ and $B$ according to:
   \begin{eqnarray}
   \sigma_{DP} \equiv m\frac{\sigma^{A} \sigma^{B}}{\sigma_{\rm eff} }.
   \label{eq:sigma_dps}
   \end{eqnarray} 
  The factor $m$ has a Poissonian nature \cite{TH3} and should be
  equal to $1/2$
  for two indistinguishable processes (like two dijet productions in $A$ and $B$)
  or gives unity for distinguishable processes. 
   The CDF  \cite{CDF97} and D0 \cite{D0_2010} experiments obtained the most accurate results on $\sigma_{\rm eff}$ 
   with an average value of about $\sigma_{\rm eff}^{\rm ave} = 15.5$ mb.

   In addition to information about parton spatial structure, those
   studies also pointed out that 
   the DP interactions can be a noticeable background to many rare processes,
   especially for those with multijet final state. In this case an additional partonic interaction,
   producingmost likely a dijet final state, can mimic the multijet signal signature.
   Some estimates of the DP background to the Higgs boson production processes at the LHC
   have been done in \cite{WH,WH1,Huss,Berger_dp}.

\begin{figure}[htbp]
\begin{center}
\vskip -4mm
\includegraphics[scale=0.27]{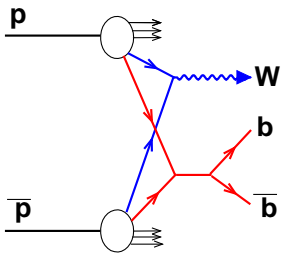}
\vskip -14mm
\caption{A possible diagram for $W+b\bar{b}$ production due to DP scattering.}
\label{fig:wbb}
\end{center}
\end{figure}

   In this paper we consider the DP events, caused by the $W$+dijet production, as a background
   to the $HW$ production, with $W\to l\nu$ and $H\to b\bar{b}$ decays, 
   which is one of the most promising Higgs boson search channels at the Tevatron.
   An example of a possible DP process with $W+b\bar{b}$ production is shown in figure \ref{fig:wbb}.
   However, in addition to the two-$b$-jet final state produced in the second parton scattering,
   we also expect significant contribution from final states with light+heavy flavor 
   and two light jets.

Due to the similarity of $HW$ and $HZ$ final states, we expect that the relative DP background
from $Z$+dijet production to the $HZ$ events should be quite close to the $HW$ case.
For this reason, we limit our study to DP background to the $HW$ events only.

   This paper is organized as follows.
   In section \ref{Sec:simsel} we describe how  DP and Higgs boson 
   samples are simulated and selected.
   In section \ref{Sec:crosssec} we calculate differential cross sections
   $d\sigma/dM_{jj}$ (where $M_{jj}$ is the invariant mass of the two leading jets) 
   and event yields in the $HW$ and DP processes including 
   the jet energy detector smearing and $b$-jet identification effects.
   The rates of events with $W+$2-jet production due to the DP and conventional single parton (SP)
   scatterings are compared in section \ref{Sec:dp_sp}.
   In section \ref{Sec:ann_dp} we introduce a set of variables
   sensitive to the kinematics of  the signal $HW(Z)$ and DP background final states
   and use them as an input to a dedicated Artificial Neural Network (ANN)
   to separate the two event types.
   We make our conclusions in section \ref{Sec:conclusion}.

\section{Simulation and selections}
\markright{\thesubsection\hspace{1em}Simulation and selection of signal and background event}{} 
\label{Sec:simsel}
\subsection{Selections}

%
%

The current {\sc pythia} event generator \cite{PYT} is the best framework to study many effects related 
to MPI production. It includes a few sophisticated phenomenological models
which consider the MPI scatterings with their various correlations, including parton momentum and color.
The MPI models in {\sc pythia}~6, have been tuned to experimental results,
and reproduce many observables in data 
quite well \cite{TH3,Sjost}. 
{\sc pythia}~8, which inherited the majority of features of its predecessor,
also allows the combination of different kinds of parton processes
in the first (main) and second scatterings within kinematic regions of interest.
%
%
To simulate events for the study we used {\sc pythia}~8 with Tune 2C as an MPI model\footnote{This tune was suggested by the {\sc pythia} authors.}.
The $HW$ production channel simulated with
Higgs boson masses of $m_H=115$ and $150$ GeV was considered. 
The DP scattering was simulated as inclusive $q\bar{q}\to W+X$ 
production in the first parton process  and
inclusive QCD dijet production in the second process.
To increase statistics in the selected final states with the cuts above,
the $W$ scattering process is required to have invariant mass 
$50<m_W<120$ GeV and the minimal allowed parton transverse momentum ($\pth$) in 
the dijet process is required to be $\pth=10$ GeV.

The event selection criteria are taken from \cite{WH_plb} and applied to both, the $HW$ and DP
production events and briefly summarized below: \\
$\bullet$ The Higgs boson is required to decay into $b\bar{b}$. \\
$\bullet$ The W-boson is selected in the electron and muon decay modes
with lepton $p_T>15$ GeV and pseudorapidity 
$|\eta|<1.1$ or $1.5<|\eta|<2.5$ for electrons and $|\eta|<1.6$ for muons.\\
$\bullet$ The total vector sum $\vec{p}_T$ of neutrinos should be $>20$ GeV (an approximate analog of missing $E_T>20$ GeV in \cite{WH_plb}). \\
$\bullet$ At least two jets are required with $p_T>20$ GeV and $|\eta|<2.5$.
  Jets are found by the D0 Run~II midpoint cone algorithm with radius $R$=0.5 \cite{blazey}.
  For this aim we used the {\sc fastjet} package \cite{fastjet} interfaced to {\sc pythia} 8.\\
$\bullet$ The scalar sum of the jet transverse momenta ($HT$) is required to be 
   $HT>60$ GeV for the 2 jet final state and $HT>80$ GeV for the 3 jet one.


\subsection{Normalizations}

The cross sections of the simulated events 
were normalized to either
experimentally measured cross sections or to theoretical NNLO predictions.
Specifically, we normalized all the {\sc pythia} cross sections in the following way:\\
$\bullet$ We simulated dijet events production 
   and calculated 
   cross sections in the dijet mass bins $150-175$ and $175 - 200$ GeV,
   and the two rapidity regions of $|y|<0.4$ and $0.4<|y|<0.8$  
   available from the recent D0 measurement \cite{dijet}. We have found
   that a required {\sc pythia}-to-data correction factor (``K-factor'')
   is about 1.26, approximately valid for both the dijet mass bins and the two rapidity regions. \\
$\bullet$ We also simulated separately $W$ inclusive production and, from a comparison of its
   cross section with the D0 and CDF measurements \cite{WZ_meas, WZ_meas1}, have obtained a {\sc pythia}-to-data 
   K-factor of about 1.5.\\
$\bullet$ The $HW$ cross section has been normalized to the NNLO predictions 
   \cite{HW_nnlo}  with the {\sc pythia}-to-NNLO K-factor equal to 1.45.\\
$\bullet$ We corrected the effective cross section $\sigma_{\rm eff}$
   used in Tune 2C 
\footnote{The effective cross section $\sigma_{\rm eff}$ in {\sc pythia} 8  is taken 
as a ratio of a total non-diffractive cross section to an impact-parameter enhancement factor,
depending on the parton spatial density distribution.}
by a factor 1.6 to match the CDF and D0 measurements 
\cite{CDF97,D0_2010} with averaged result  $\sigma_{\rm eff}^{\rm ave}$  = 15.5 mb.

   The uncertainty assigned in our analysis to the K-factors are 10\% 
   and 16\% to $\sigma_{\rm eff}^{\rm ave}$. The latter is due to the  
   difference between the D0 and CDF $\sigma_{\rm eff}$ central values ($\sim\!7\%$) 
   and the systematic uncertainties ($\sim\!14\%$) in the D0 measurement.

\section{$d\sigma/dM_{jj}$ cross sections for HW and double parton events}
\markright{\thesubsection\hspace{1em}$d\sigma/dM_{jj}$ cross sections for HW, HZ and double parton events}{} 
\label{Sec:crosssec}

\subsection{HW and DP cross sections}

\begin{figure}[htbp]
\hspace*{-2mm} \includegraphics[scale=0.4]{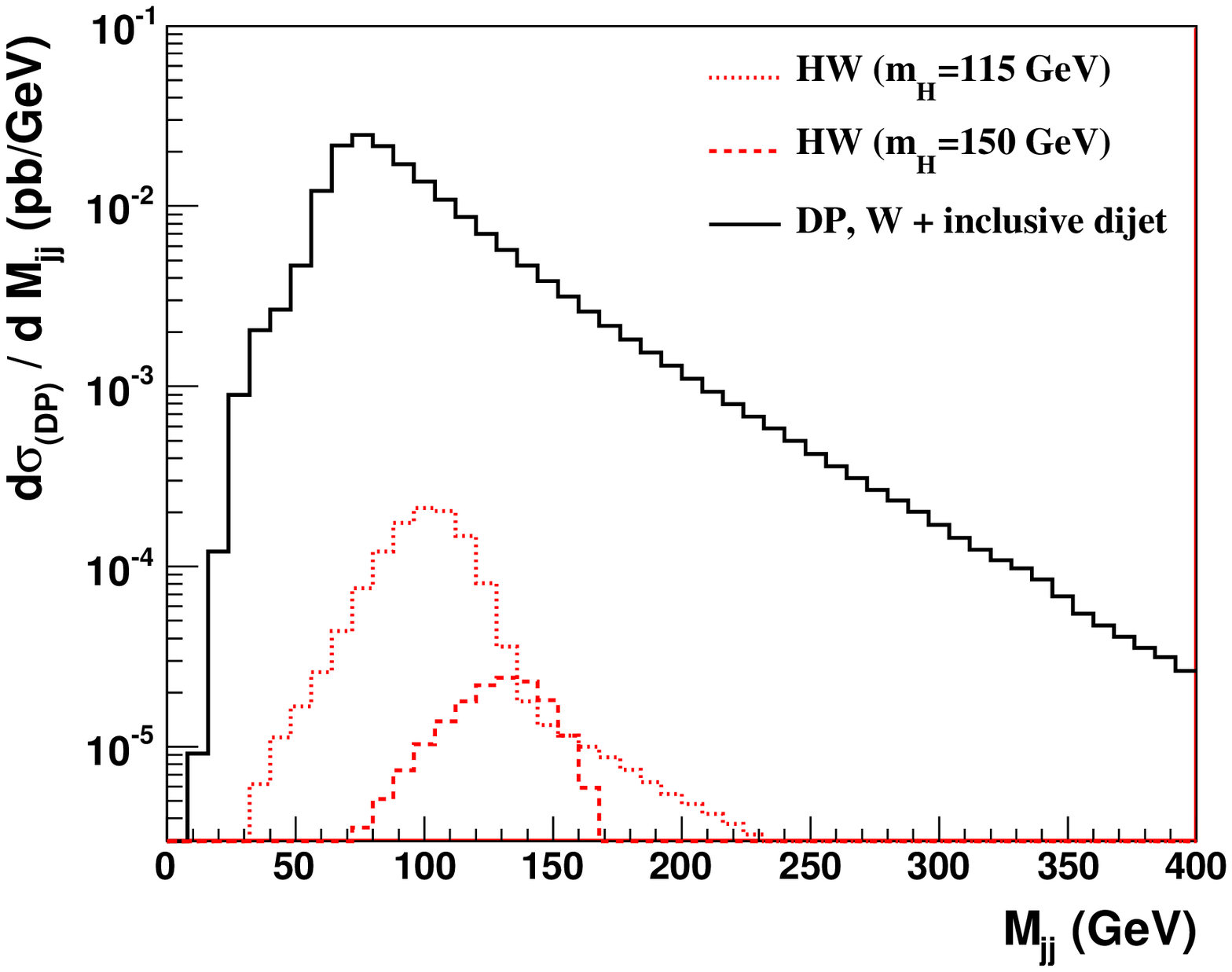}
\hspace*{-2mm} \includegraphics[scale=0.4]{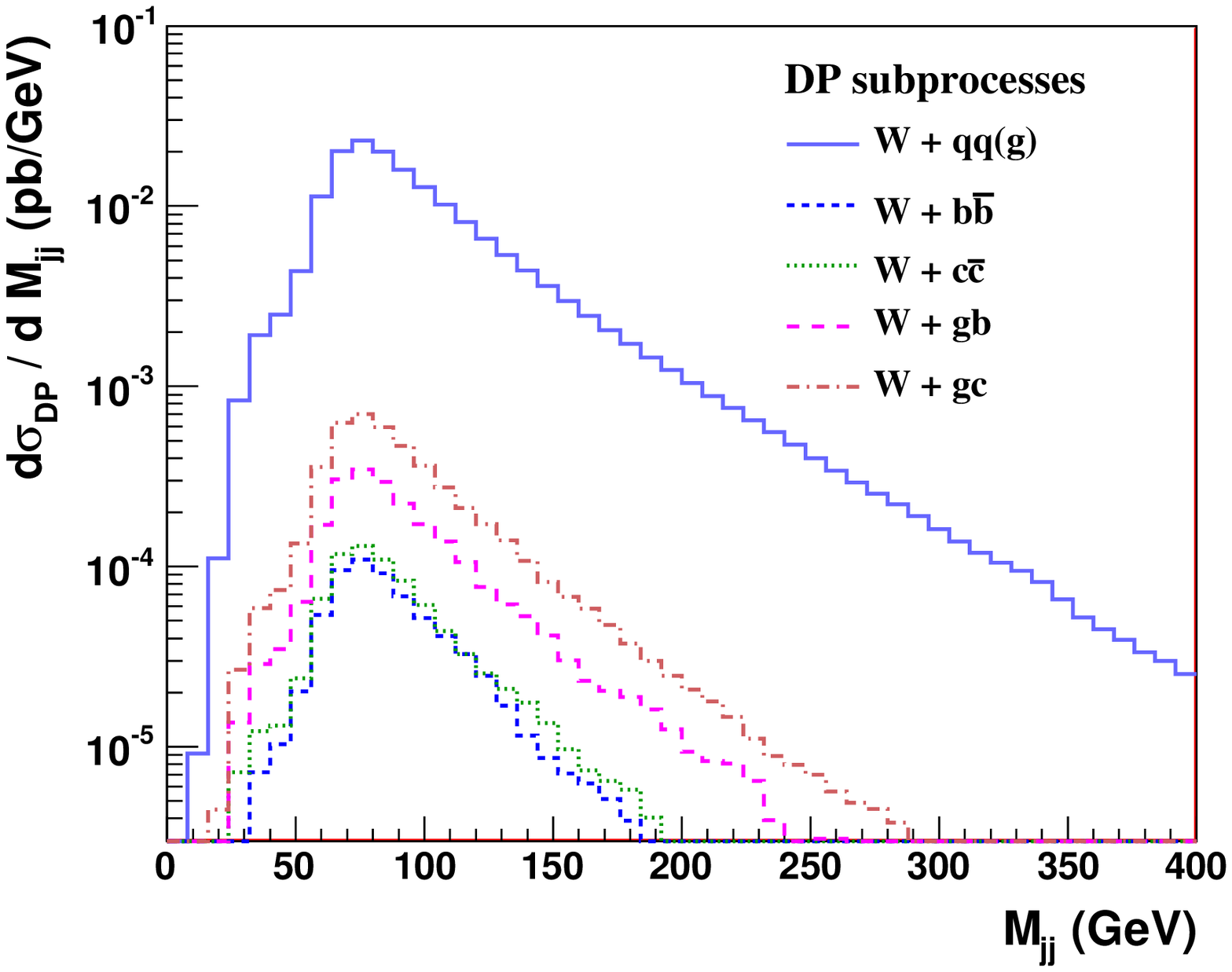}
\vskip -4mm
\caption{The differential cross sections in the dijet mass $M_{jj}$ bins for 
signal $HW$ and background DP events including the jet $p_T$ resolution. 
On the left plot, dotted and dash-dotted red lines correspond to $HW$ events with $m(H)=115$ and $150$ GeV,respectively, while the
full black line shows the total background from all the DP $W$+dijet channels.
The right plot shows contributions from main parton scattering subprocesses 
composing the total DP background.
}
\label{fig:XS_hw_sm}
\end{figure}

In this section we calculate he differential cross sections $d\sigma/dM_{jj}$
for the $HW$ and DP ($W$+dijet) events selected according to the criteria of section \ref{Sec:simsel}.
To match the detector resolution, the jet transverse momenta $p_T$ are smeared using
\begin{eqnarray}
\frac{\sigma_{p_T}}{p_T} = \frac{S}{\sqrt{p_T}} \oplus C,
\label{eq:resol}
\end{eqnarray}
where $S=0.75$ and $C=0.06$ which approximately reproduces the jet $p_T$ resolution for
the D0 detector \cite{Mikka}.
The differential cross sections $d\sigma/dM_{jj}$ for the $HW$ and DP productions including the smearing effect 
are shown in figure \ref{fig:XS_hw_sm}. In addition to the total DP cross section,
contributions from the main DP scattering subprocesses are also shown in a separate plot.
One can see from these two plots that (a) the DP cross section dominates the $HW$ signal 
by more than two orders of magnitude, and (b) the DP cross section is caused mainly
by the $W$+2 light jets (stemming from $u/d/s$-quarks or gluons) production, followed, 
in the order of importance, by contributions from 
$W+gc$, $W+gb$, and then by $W+c\bar{c}$ and $W+b\bar{b}$ events.



\subsection{Account of $b$-jet identification efficiencies}

\begin{figure}[htbp]
\hspace*{-2mm} \includegraphics[scale=0.4]{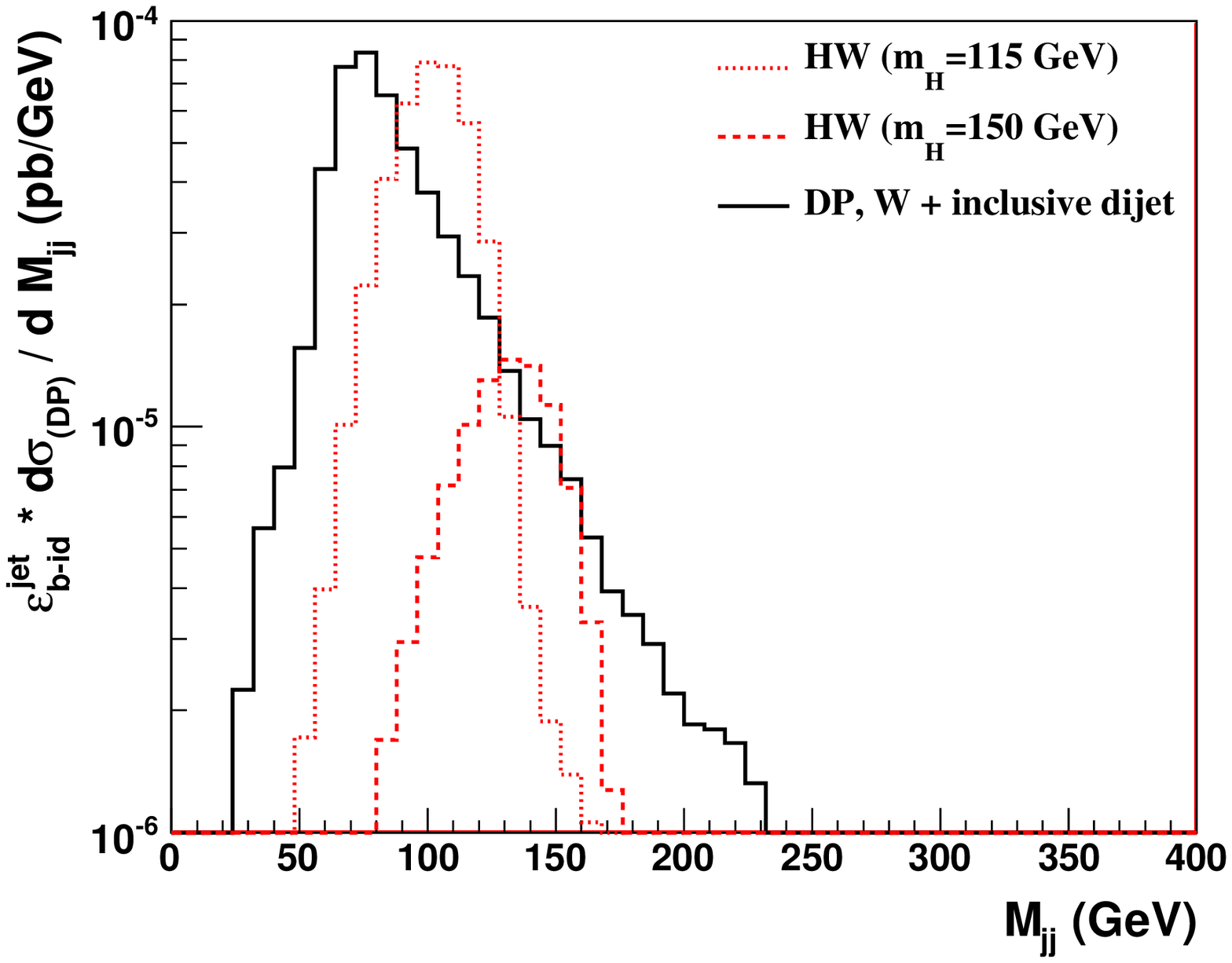}
\hspace*{-2mm} \includegraphics[scale=0.4]{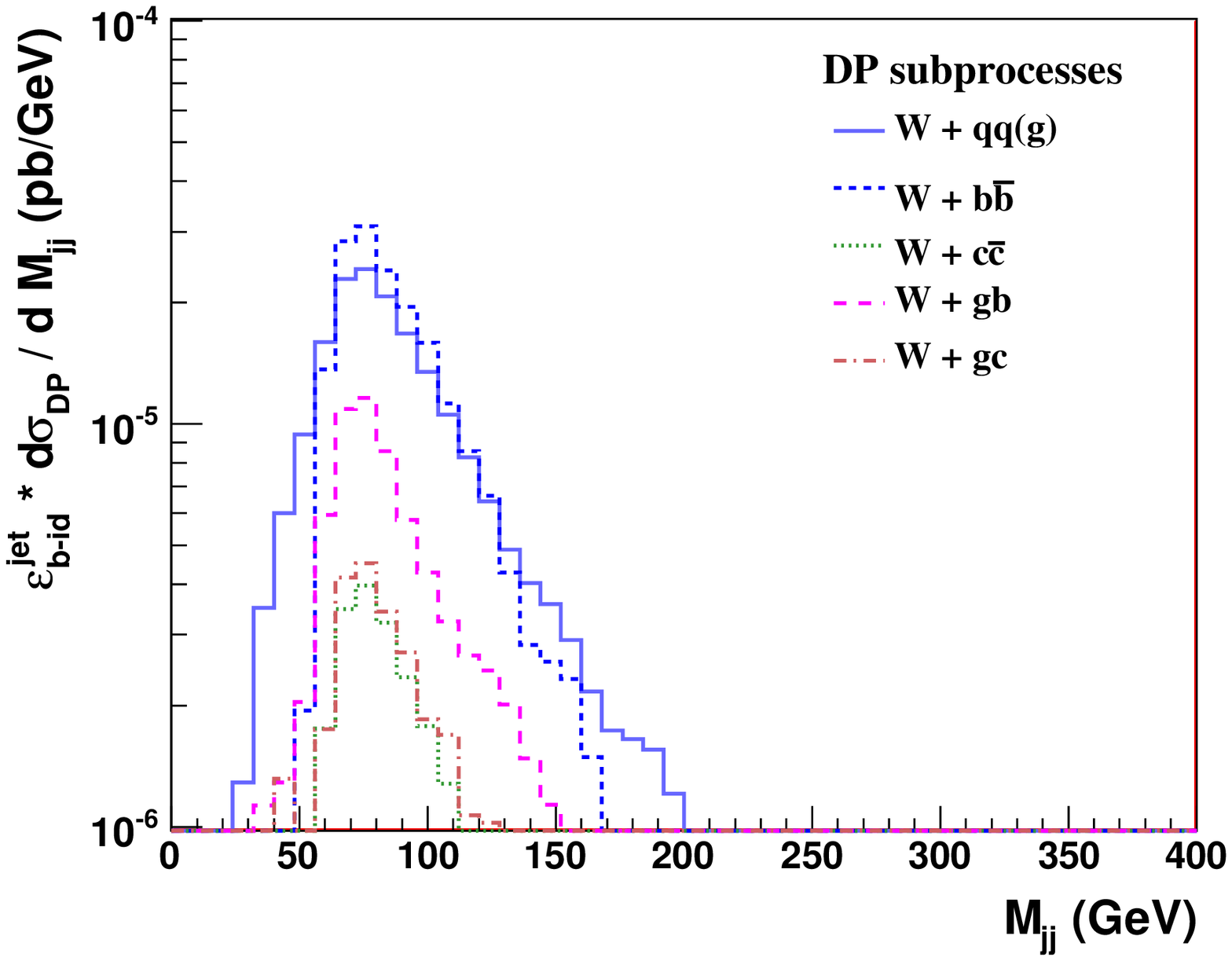}
\vskip -4mm
\caption{The differential cross sections in the dijet mass bins for signal $HW$ and background DP events 
including the jet $p_T$ resolution and requirement of the two jet $b$-tagging
(See also description in the caption to figure \ref{fig:XS_hw_sm}).
}
\label{fig:XS_hw_sm_dt}
\end{figure}

\begin{figure}[htbp]
\hspace*{-2mm} \includegraphics[scale=0.4]{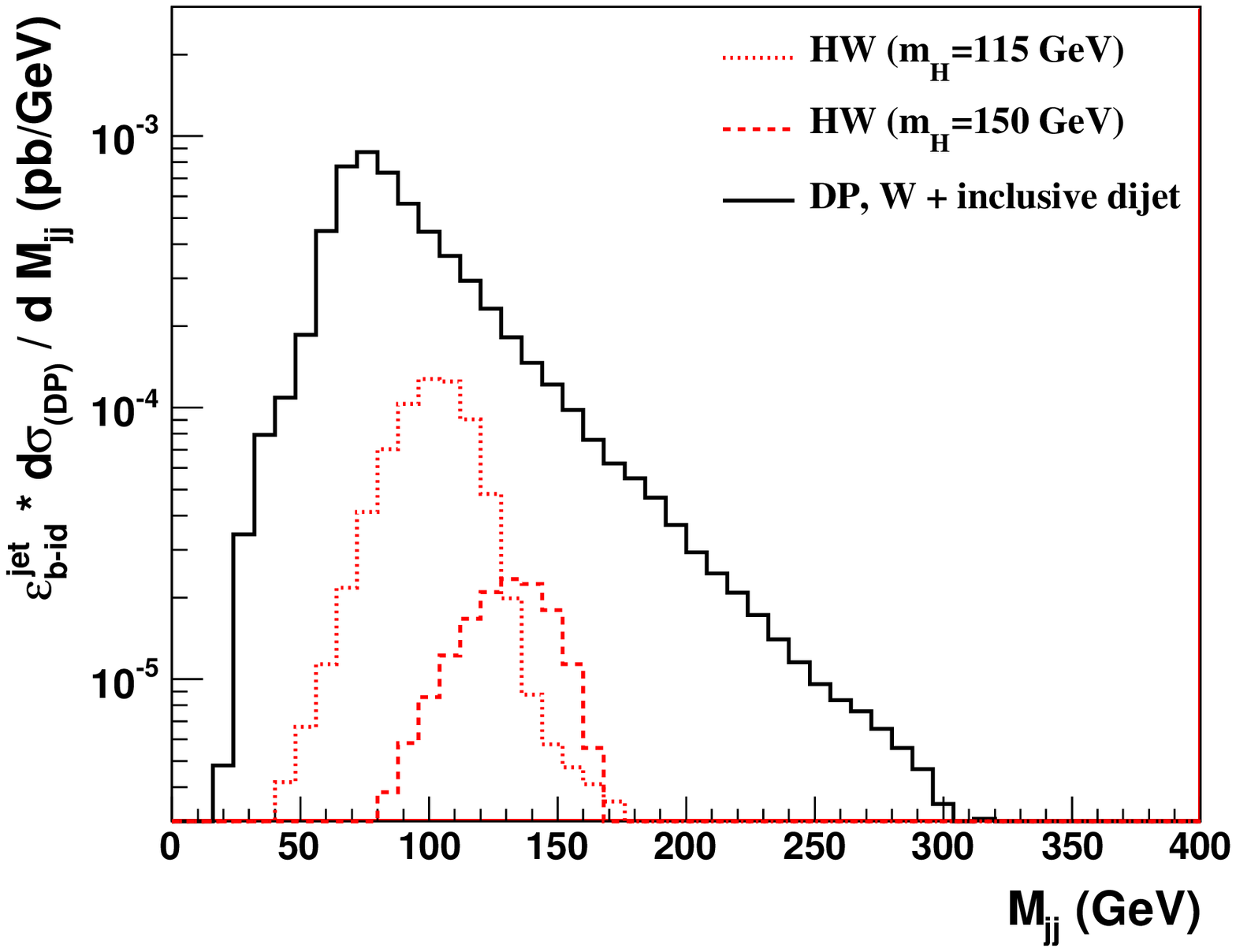}
\hspace*{-2mm} \includegraphics[scale=0.4]{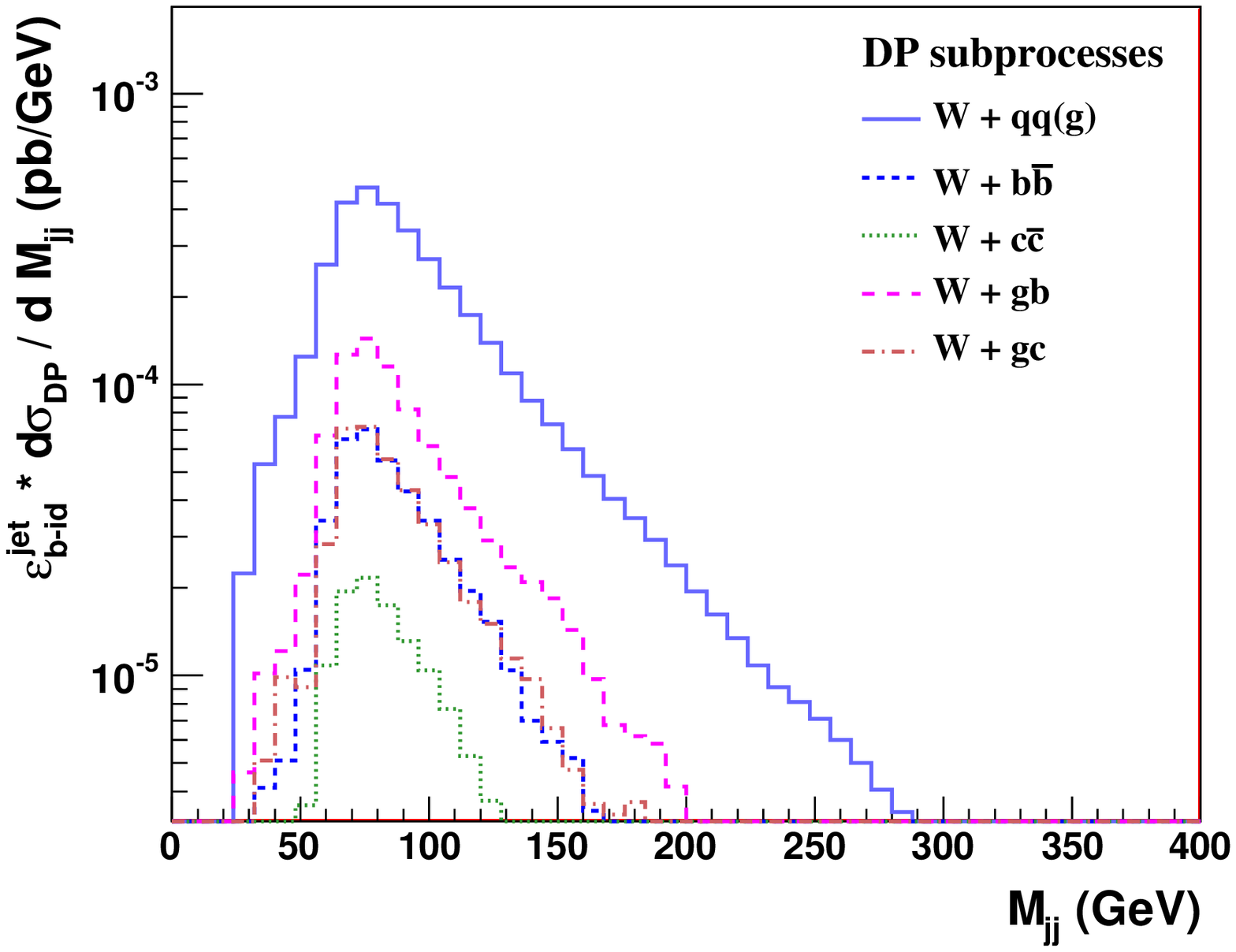}
\vskip -4mm
\caption{The differential cross sections in the dijet mass bins for signal $HW$ and background DP events 
including the jet $p_T$ resolution and the combined jet $b$-tagging efficiency (see also the main text and 
the caption to figure \ref{fig:XS_hw_sm}).}
\label{fig:XS_hw_sm_comb}
\end{figure}

In the signal $HW$ events we have two $b$ jets in the final state. Since the leading DP background
is caused by the $W+$2 light jet events (figure \ref{fig:XS_hw_sm}),
we should expect a significant reduction after requiring of jet
$b$-tagging. To check this numerically, we apply
a specific $b$-tagging requirement for the $HW$ and DP events.
In our fast MC we cannot check the jet $b$-tagging quality, but we instead use
the efficiencies to pass the $b$-tagging requirements for light ($l$), $c$ and $b$ jets.
We take these efficiencies from \cite{trf}, where they are
parametrized as functions of jet $p_T$ and $\eta$. These efficiencies 
are used to re-weight events according to the jet flavors.
Typical efficiencies are $50-70\%$ for $b$-jets, $8-12\%$ for $c$-jets
and $0.5-2\%$ for $l$-jets. The variations reflect dependence on the jet $p_T, \eta$
and tightness of the $b$-tagging condition.
We consider a given jet to be a $b$-jet if it has a $b$-quark in the jet cone;
if the jet does not have a $b$-quark but has a $c$-quark instead, it is considered to be 
a $c$-jet; otherwise it is a light jet.
Figure \ref{fig:XS_hw_sm_dt} shows the cross sections $\times$ $b$-jet identification efficiency 
($\varepsilon_{\rm b-id}^{\rm jet}$) for the DP and $HW$ events,
where each of the two jets is required to satisfy
the ``loose'' $b$-tagging requirement \cite{trf}. This requirement significantly suppresses rates of the DP events. 
However, the signal rates are also noticeably reduced (compare figures \ref{fig:XS_hw_sm} and \ref{fig:XS_hw_sm_dt}).
For this reason, in practice, double tagging 
is usually combined with single tagging. 
For example, in the search for $HW$ signal \cite{WH_plb}, two cases of the $b$-tagging are considered:
either an event should contain two jets satisfying ``loose'' $b$-tagging requirements
or, if it fails, a single jet should satisfy the ``tight''  requirement.
Fractions of background (=data) and the $HW$ events 
selected with the single $b$-tagging can be taken from \cite{WH_plb}: 
they are about $85\%$ and $60\%$ 
correspondingly\footnote{Clearly, here we assume that the jet flavor content of the background 
events in data and the dijet events from the DP interaction
is the same. However, we believe that for the current level of estimates this assumption
should be good enough.}.
The remaining events are with two $b$-tagged jets.
%
Figure \ref{fig:XS_hw_sm_comb} shows the cross sections $\times \varepsilon_{\rm b-id}^{\rm jet}$
for the DP and HW events
where we have combined events with single and double $b$-tagging according to their fractions mentioned above. 
We see that while the dominating DP channel is still caused by 
the $W$+2 light jet production, the relative contribution from $W+gb$ production is now
much higher than in figure \ref{fig:XS_hw_sm} (no $b$-tagging is applied). 
The $W+gb$ contribution is followed by similar ones from the $W+gc$ and $W+b\bar{b}$ events.

\begin{figure}[htbp]
\begin{center}
\hspace*{-0mm} \includegraphics[scale=0.4]{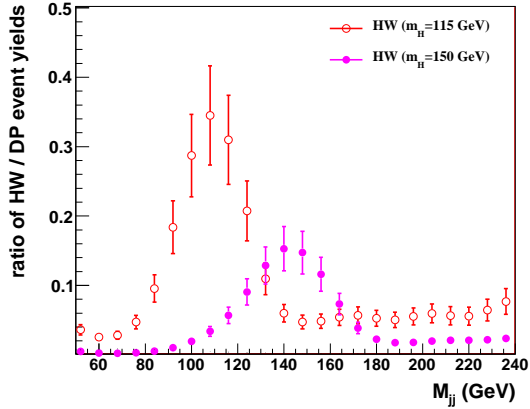}
\vskip -4mm
\caption{The ratio of $HW$ signal to DP background event yields with the combined $b$-tagging (see the main text).}
\label{fig:SB_hw_stdt}
\vskip -4mm
\end{center}
\end{figure}

Figure \ref{fig:SB_hw_stdt} is complementary to figure \ref{fig:XS_hw_sm_comb} and
shows the ratios of the $HW$ event yield to the inclusive DP $W+$dijet one in the dijet mass $M_{jj}$ bins
for the events selected by the combined 
$b$-tagging.
The uncertainty in each bin is caused by the K-factors and effective cross section (section \ref{Sec:simsel}).

%
One can see that 
the Higgs boson events with $m_{H}=115$ GeV are expected to be suppressed by about a factor 3
($S/B\simeq 0.35$) in the peak position, while the signal events with $m_{H}=150$ GeV 
are suppressed by about a factor 7.

It is interesting to compare the total number of the signal events predicted by our fast MC after all selections
(figure \ref{fig:XS_hw_sm_comb}) with those in \cite{WH_plb} for the integrated luminosity $L_{int} = 5.3~{\rm fb}^{-1}$. 
It is obtained by integrating the cross section over the whole $M_{jj}$ range (20--400 GeV)
and multiplying by $L_{int}$. In such a way we have found the expected signal statistics of about 
31 (7) events for $m_{H}=115~(150)$ GeV. 
According to \cite{WH_plb} there should be about $19\pm 1$ selected events for $m_{H}=115$ GeV.
Our estimate seems to be in a reasonable agreement if we take into account the  effects
of finite lepton identification, jet taggability efficiencies, and detector acceptance unaccounted in our fast MC.
%
%
%
%

\section{Comparison of the DP and SP event yields}
\label{Sec:dp_sp}

In this section we compare the event yields $dN/dM_{jj}$ 
expected for the DP and SP $W$+2-jet productions.
The two additional jets in the SP events  come from radiation effects in the initial and/or final states.
SP events are simulated using $q\bar{q}\to Wg$ and $qg\to Wq$ subprocesses
and applying the $HW$ selection criteria from section \ref{Sec:simsel}.
To reproduce the inclusive $W$+2 jet cross section in data \cite{CDF_Wj},
the {\sc pythia} events are reweighted with a scaling factor 
depending on the second jet $p_T$, what increases the {\sc pythia} $W$+2 jet cross section 
in the region $110 < M_{jj}< 160$ GeV by about a factor 2. 
Also, as before, the jet $p_T$ is smeared according to the $p_T$ resolution, eq. (\ref{eq:resol}) and
the events are weighted with the jet $b$-tagging efficiencies according to the jet flavors.

\begin{figure}[htbp]
\begin{center}
\hspace*{-0mm} \includegraphics[scale=0.4]{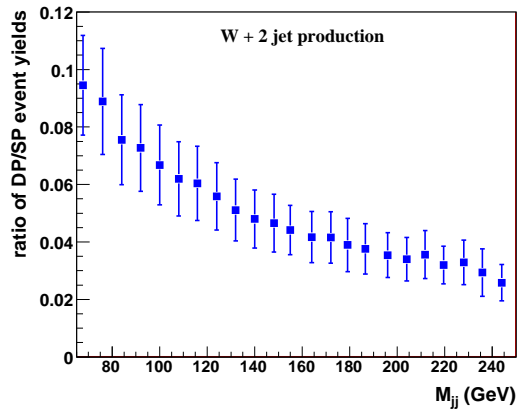}
\vskip -4mm
\caption{The ratio of the DP to SP event yields for the $W+$2-jet production.}
\label{fig:wh_spdp}
\vskip -4mm
\end{center}
\end{figure}

The estimated total event yields 
in the whole mass region at $L_{int} = 5.3~{\rm fb}^{-1}$
for SP and DP events are about 5212 and 262 events, respectively.
The differential ratios of the DP/SP $W$+2-jet event yields in the $M_{jj}$ bins are shown in 
figure \ref{fig:wh_spdp}.
They are about $5-8\%$ for $M_{jj}\simeq 115$ GeV and $3.5-6\%$ for $M_{jj}\simeq 150$ GeV.


\section{Artificial neural network for DP and HW(Z) separation}
\markright{\thesubsection\hspace{1em}Artificial neural network for DP and HW(Z) separation}{} 
\label{Sec:ann_dp}

\subsection{Variables}

\label{Sec:vars}


In this section we discuss variables that can be useful to separate
the $HW(Z)$ signal from the DP $W(Z)+$dijet background events.
Most of these variables are either based on the previous relevant experimental studies
\cite{AFS, UA2, CDF93, CDF97, E735, D003,D0_2010} or have been suggested in theoretical papers
\cite{TH3, Landsh, Goebel, TH1, TH11, TH2, TH21, Mang}.
Due to the similarity of $HW$ and $HZ$ events, most of these
variables should be useful to suppress the DP background events to both the final states
(with some exclusions).
Definitions of all the variables are summarized below.\\
\noindent
$\bullet$ The first variable is an azimuthal angle between two $p_T$ vectors, where the first one
corresponds to the $W(Z)$ $p_T$ vector, while the second one
is a sum of the leading and second jet $p_T$ vectors:
\begin{eqnarray}
\Delta S \equiv \Delta\phi\left(\vec{p}_{T}[V], ~\vec{p}_{T}[{\rm jet_1, jet_2}]\right),
\label{eq:DeltaS_var}
\end{eqnarray}
where $\vec{p}_{T}[V]$ is the transverse momentum vector of $V(=W,Z)$-boson 
and $\vec{p}_{T}[{\rm jet_1, jet_2}] = {\vec p}_T^{\rm ~jet_1} + {\vec p}_T^{\rm ~jet_2}$.
For historical reasons \cite{AFS,UA2,CDF93,CDF97,D0_2010} we call
this angle $\Delta S$.

\noindent
$\bullet$ The second variable is 
the difference between the rapidity of the $V$-boson and the total rapidity of the two-jet system:
\begin{eqnarray}
\Delta \eta({\rm V,jet12}) =  |\eta^{\rm V} -  (\eta^{\rm jet1} + \eta^{\rm jet2})|.
\label{eq:DeltaEta}
\end{eqnarray}
%

%
\noindent
$\bullet$ The variable $\Delta\eta({\rm V,jet12})$ 
can be calculated just for $V=Z$ events, but not for $W$ 
due to the missing $p_z$ information of the $\nu$. Instead we can use the rapidity 
of the electron ($e$) $\eta^e$ from the $W$ decay and introduce an analogous variable:
\begin{eqnarray}
\Delta \eta({\rm e,jet12}) =  |\eta^{e} -  (\eta^{\rm jet1} + \eta^{\rm jet2})|.
\label{eq:DeltaEta}
\end{eqnarray}
%
\noindent
$\bullet$ In the case of the $W$ production the azimuthal angle between the electron from the $W$ decay 
and the leading jet $\Delta \phi({\rm e,jet1})$ can be considered.

Two other variables use angular differences between the first and second jets:\\
\noindent
$\bullet$ the azimuthal angle between the  jets $\Delta \phi({\rm jet1, jet2})$.\\
\noindent
$\bullet$ the difference between rapidities of the first and second jets $\Delta \eta({\rm jet1, jet2})$.\\
\noindent
$\bullet$ Another variable characterizes the orientation of the two event planes, 
one contains the beam (proton) axis and $V$-boson, and the other one contains
the two jets \cite{RunI_34jets}:
\begin{eqnarray}
{\rm cos}\psi^\star({\rm V,jet12}) =  \frac{(\vec{p}^{~V}\times\vec{p}^{\rm ~proton}) \cdot (\vec{p}^{\rm ~jet1}\times\vec{p}^{\rm ~jet2})}
{|\vec{p}^{~V}\times\vec{p}^{\rm ~proton}| \cdot |\vec{p}^{\rm ~jet1}\times\vec{p}^{\rm ~jet2}|}.
\label{eq:CosPsi}
\end{eqnarray}
%
%
\noindent
$\bullet$ In the case of the $W$ production, we do not have the 3-vector of the $W$ momentum 
but can use the electron  3-vector instead,
i.e. we should calculate ${\rm cos}\psi^\star(e,{\rm jet12})$.

Three other variables are based on the jet $p_T$:\\
\noindent
$\bullet$ the total sum of the first and second jet $p_T$:
\begin{eqnarray}
p_{T}^{\rm sum12} = p_{T}^{\rm jet1} + p_{T}^{\rm jet2}.
\label{eq:PTsum}
\end{eqnarray}
\noindent
$\bullet$ the relative difference between the first and second jet $p_T$:
\begin{eqnarray}
p_{T}^{\rm diff12} = (p_{T}^{\rm jet1} - p_{T}^{\rm jet2}) / p_{T}^{\rm sum12}. 
\label{eq:PTdiff}
\end{eqnarray}
\noindent
$\bullet$ the total $p_T$ sum of all jets, $p_{T}^{\rm sumAll}$.\\
\noindent
$\bullet$ Finally, we add the total number of all jets ($p_T>6$ GeV), $N_{\rm jets}$.

\begin{figure}[htbp]
\begin{center}
\hspace*{-0mm} \includegraphics[scale=0.7]{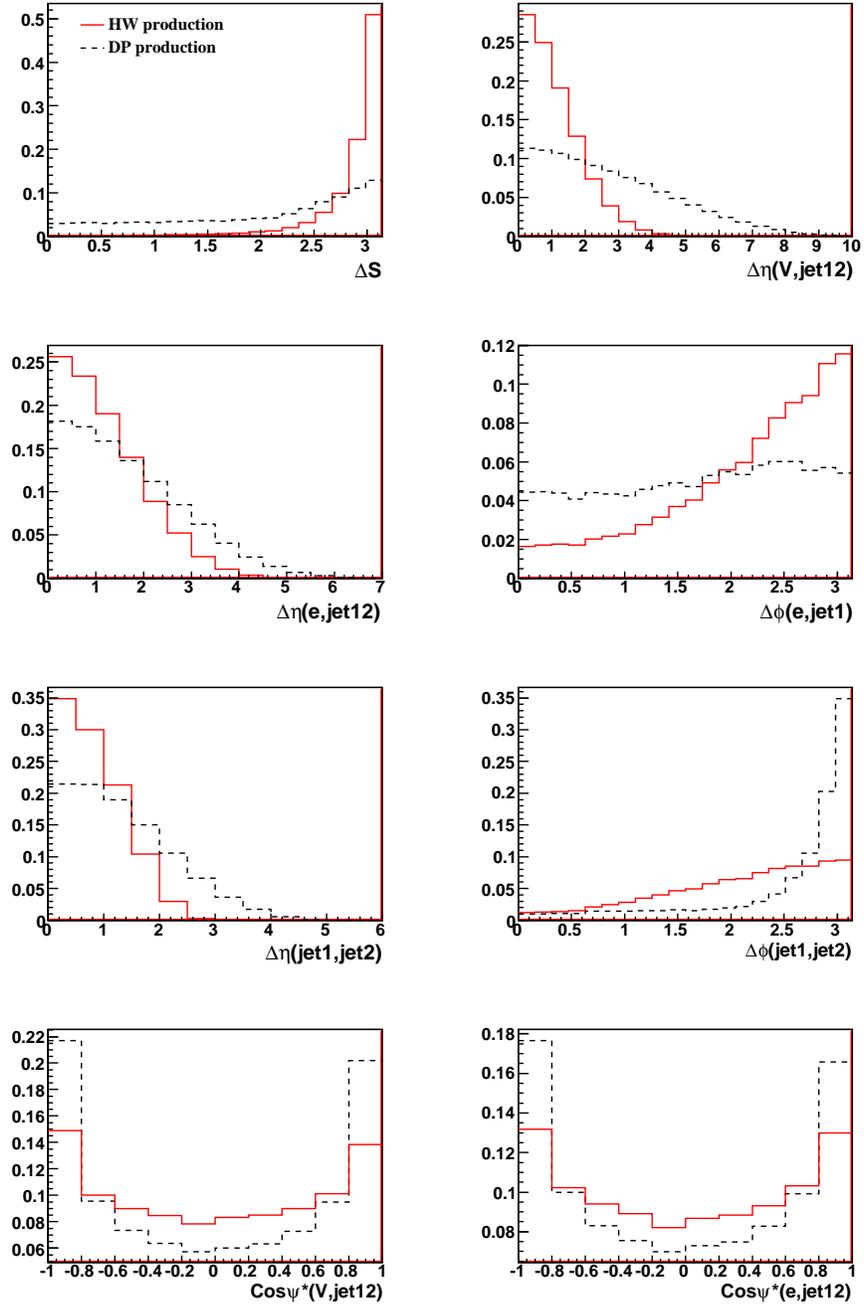}
\caption{Normalized distributions of the number of $HW$ signal (full red line) and $W+$dijets background 
(dashed black line) events over the kinematic variables of section \ref{Sec:vars} (part 1).}
\label{fig:ann1}
\end{center}
\end{figure}
\begin{figure}[htbp]
\begin{center}
\hspace*{-0mm} \includegraphics[scale=0.65]{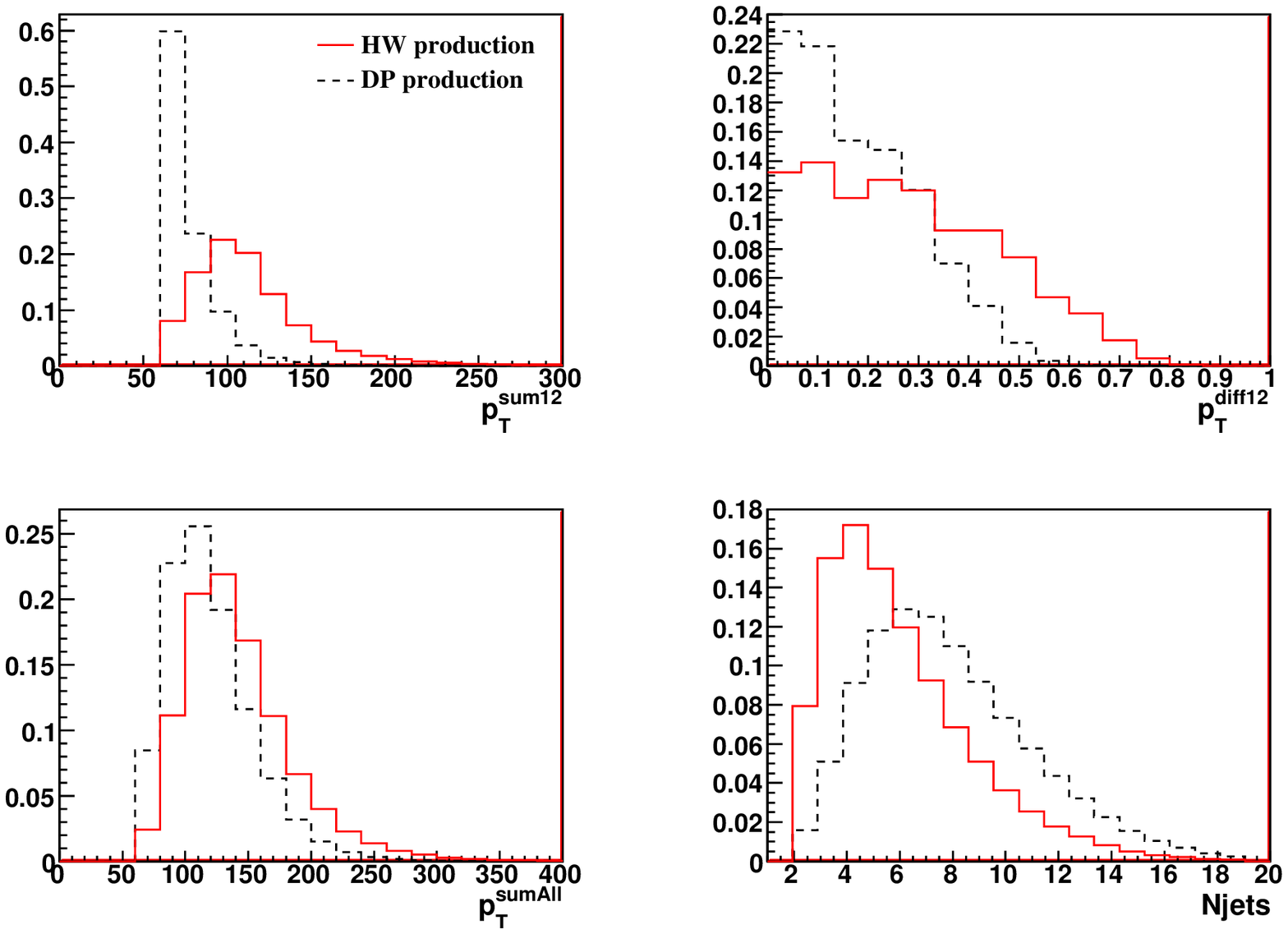}
\caption{Normalized distributions of the number of $HW$ signal (full red line) and $W+$dijets background 
(dashed black line) events over the kinematic variables of section \ref{Sec:vars} (part 2).}
\label{fig:ann2}
\end{center}
\end{figure}

All these 12 variables are shown in figures ~\ref{fig:ann1} and \ref{fig:ann2} for $HW$ and DP $W+$dijet
events. They demonstrate a good separation power between the two event types.


\subsection{ANN}

The variables presented above can be used as input to a dedicated ANN 
to separate the $HW$ from the DP events.
The variable $p_{T}^{\rm sumAll}$ is very correlated with $p_{T}^{\rm sum12}$,  
but the latter is a bit more sensitive to the signal/background difference.
We do not use the dijet mass information to minimize dependence on a specific Higgs boson
mass region but rather concentrate on other more generic kinematic properties of the two event types.

\begin{figure}[htbp]
\begin{center}
\hspace*{-0mm} \includegraphics[scale=0.39]{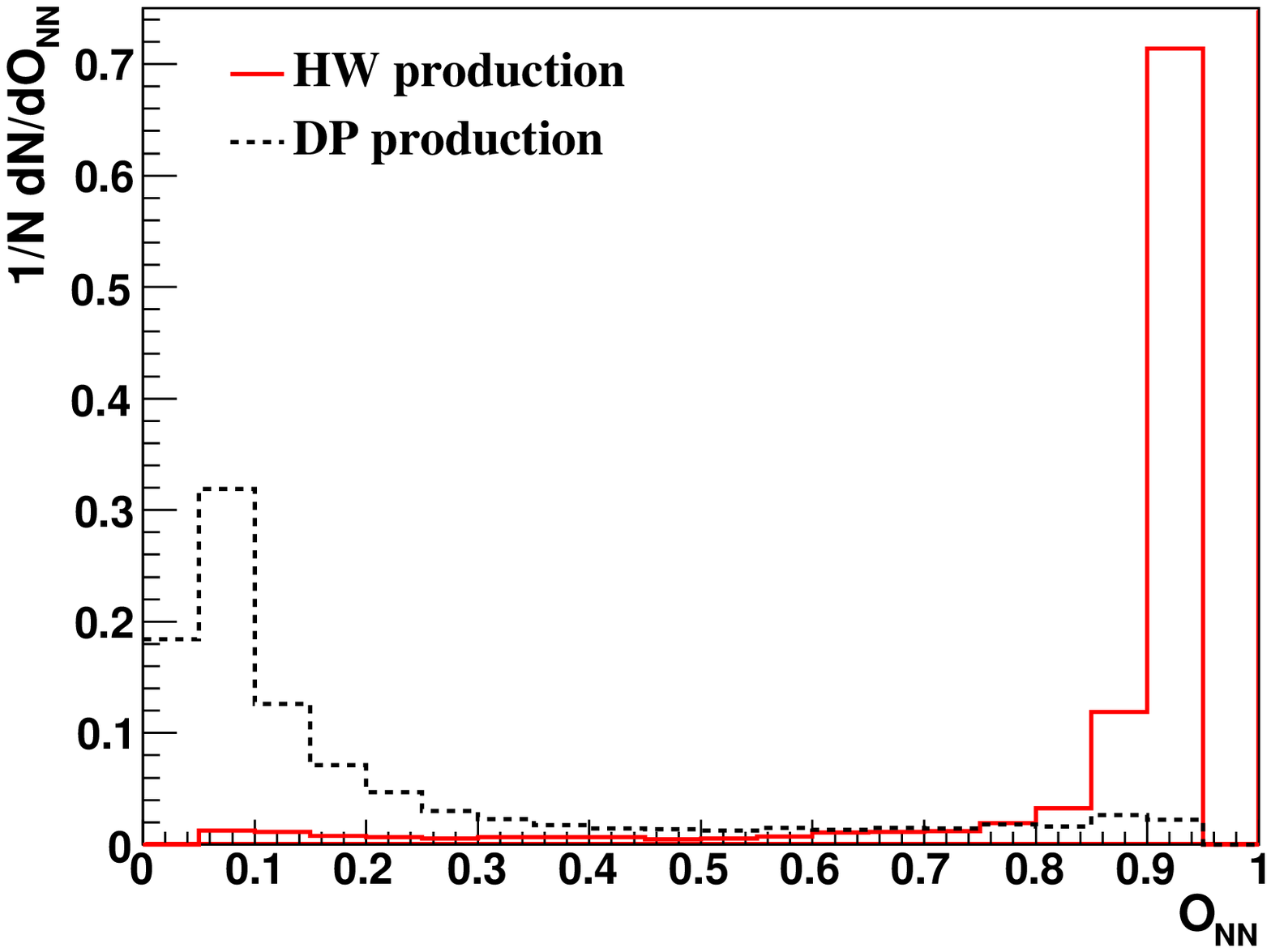}
\vskip-4mm
\caption{The ANN output for the DP and $HW$ ($m_H=115$ GeV) events using the 9 input variables described in the text.}
\label{fig:nnout}
\end{center}
\end{figure}

%
We have chosen the following 9 variables to train the ANN:
$\Delta S$, $\Delta \eta({\rm e, jet12})$, $\Delta \phi({\rm e,jet1})$, $\Delta \phi({\rm jet1, jet2})$, 
$\Delta \eta({\rm jet1, jet2})$, ${\rm cos}\psi^\star(e,jet12)$, $p_{T}^{\rm sum12}$, $p_{T}^{\rm diff12}$,  
and $N_{\rm jets}$, using the package {\sc jetnet} \cite{JN}. 
The ANN is trained using the signal $HW$ (simulated with $m_{H}=115$ GeV) and background DP events 
to produce a single output value equal to zero for the background and unity for the signal events. 
The DP background events for the training (and later for testing) purposes are selected around 
the Higgs boson $M_{jj}$ peak position taking all events within $\pm 2\sigma$ around the peak.
We have trained the ANN using 200,000 the signal and background events and then tested the ANN
using 50,000 events that have not been used at the training stage.
The normalized distributions of the signal and background events 
for the ANN output $O_{\rm NN}$ is presented in figure \ref{fig:nnout}.
The ANN weights obtained at the training stage, 
have been used later to also separate the $HW$ signal simulated with $m_{H}=150$ GeV and DP events.

\begin{figure}[htbp]
\begin{center}
\hspace*{-0mm} \includegraphics[scale=0.39]{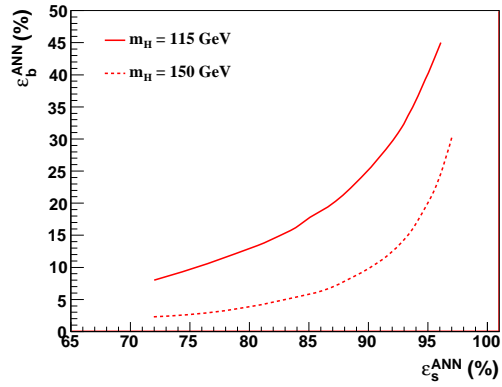}
\vskip-4mm
\caption{DP versus $HW$ neural network selection efficiencies.}
\label{fig:anneff}
\end{center}
\end{figure}

Tighter cuts on the ANN output will reject a larger fraction of the DP events.
Figure \ref{fig:anneff} shows the correlation between efficiencies to select the background and signal events 
($\varepsilon_{b}^{\rm ANN}$ and $\varepsilon_{s}^{\rm ANN}$, respectively)
for the two Higgs boson masses, $m_{H}=115$ GeV and $m_{H}=150$ GeV. 
Selecting $90\%~(80\%)$ of the signal events with $m_{H}=115$ GeV we will keep only 
about $24\%~(13\%)$ of the DP events,
while selecting $90\%~(80\%)$ of the signal events with $m_{H}=150$ GeV 
we will keep about $9\%~(4\%)$ of the DP events.

\subsection{Results}

\begin{figure}[htbp]
\hspace*{-2mm}\includegraphics[scale=0.4]{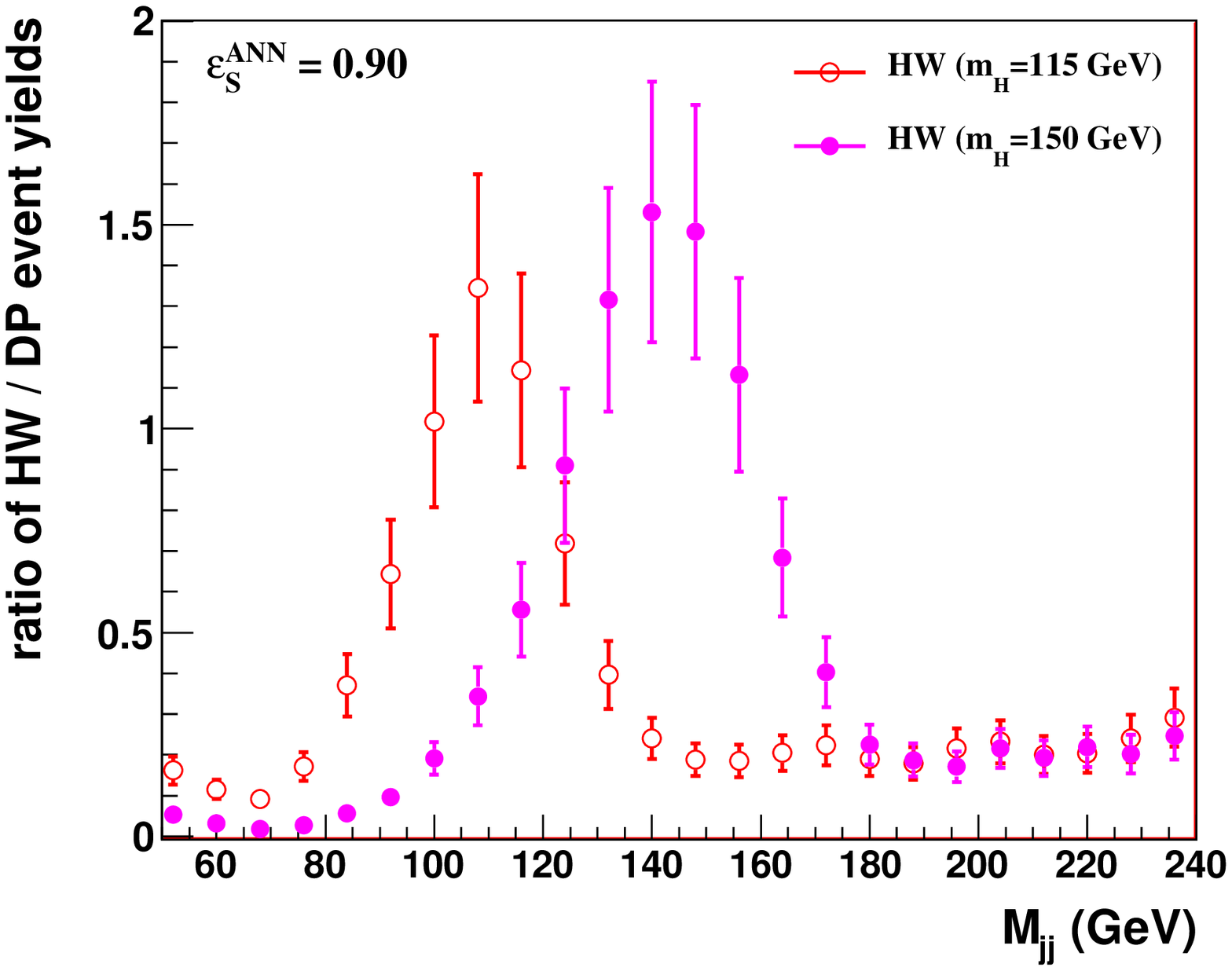}
\hspace*{-2mm} \includegraphics[scale=0.4]{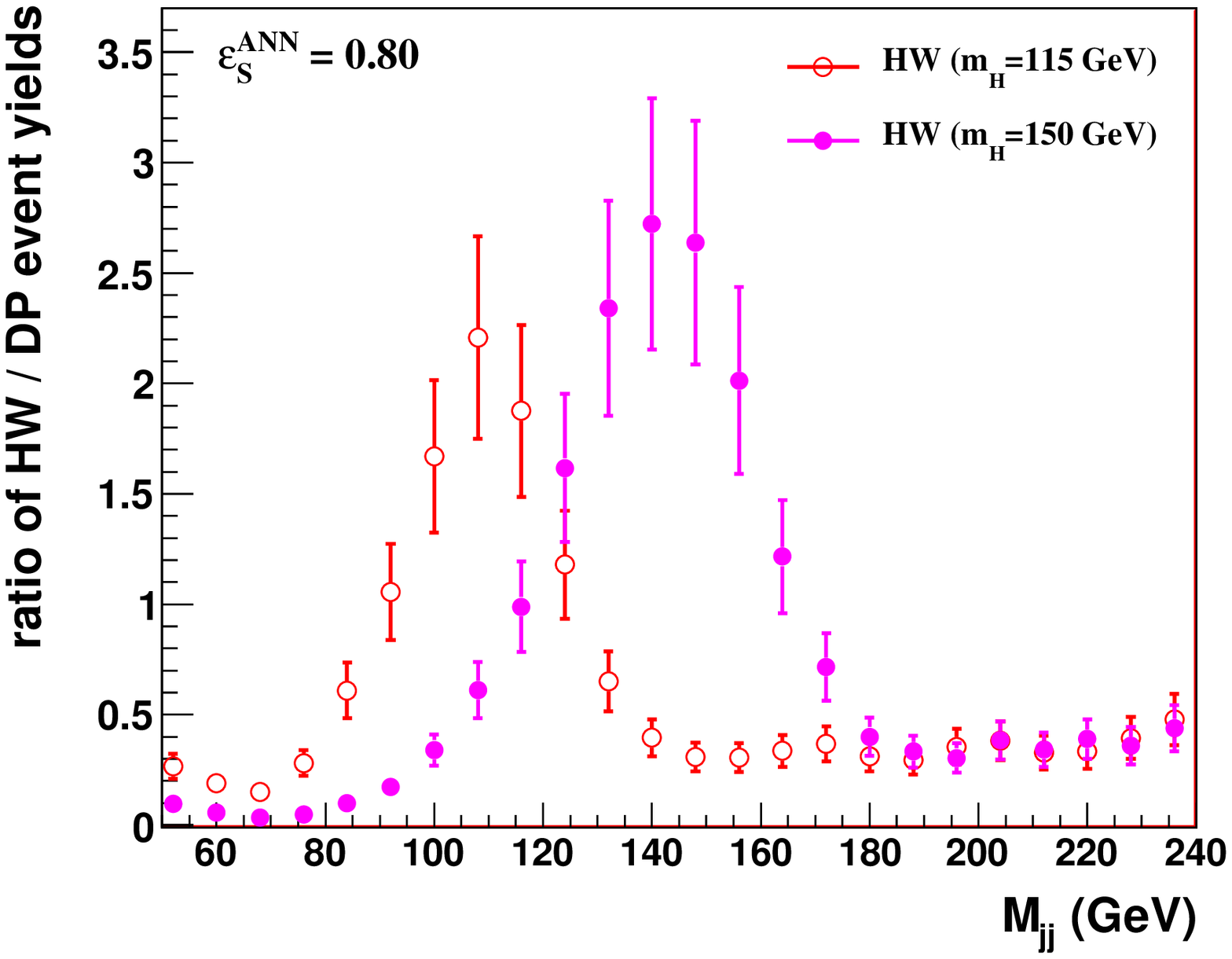}
\vskip-3mm
\caption{Ratio of the $HW$ event yields to the DP ones with account of the ANN selection efficiencies taken 
for the $HW$ events to be 90\% on the left and 80\% on the right plot.}
\label{fig:SB_hw_stdt_8590}
\end{figure}

The built ANN is used to further suppress the DP background, which strongly dominates the signal events
even after the b-tagging selections (figure \ref{fig:SB_hw_stdt}). 
The new signal-to-background ratios are shown in two plots of figure \ref{fig:SB_hw_stdt_8590}, corresponding
to the choice of the $HW$ signal efficiencies $\varepsilon_{s}^{\rm ANN} = 90\%$ and $80\%$.
The ratios at $\varepsilon_{s}^{\rm ANN} = 90\%$ for both the mass regions, 115 GeV and 150 GeV,  
are now close to $1.3-1.5$. This ratio grows further with  $\varepsilon_{s}^{\rm ANN} = 80\%$, 
and reaches about 2.2 at $M_{jj}\simeq 115$ GeV and about 2.7 at $M_{jj}\simeq 150$ GeV.

\section{Conclusion}\markright{\thesubsection\hspace{1em}Conclusion}{} 
\label{Sec:conclusion}

In our current study we have shown that the $W+$dijet events produced due to the DP scattering
can compose a quite sizable background to the associated $HW$ production with $H\to b\bar{b}$ decay.
Its relative fraction with respect to the traditional background from SP scattering
with the $W+$2-jet final state is found to be $4-8\%$ in the dijet mass region $115<M_{jj}<150$ GeV.
We suggest a set of the angular and jet $p_T$ variables that are sensitive to the difference between the $HW$
and DP kinematics. The neural network built using these variables allows significant
suppression of the DP background to a desirable level. 
Provided that the overall systematics in the Higgs searches in the $HW$ channel will go down in a time
and since every percent of the background events matters, use of the suggested
anti-DP neural network should be very helpful.

~\\[2mm]
\noindent
\centerline{\bf Acknowledgments}\\[1mm]
  The authors thank Wade Fisher and Aurelio Juste for useful discussions
  and Stephen Mrena for the help with use of the {\sc pythia} 8 code. 

\clearpage
\providecommand{\href}[2]{#2}\begingroup\raggedright\endgroup


\begin{thebibliography}{10}

\bibitem{AFS}
{\bf Axial Field Spectrometer} Collaboration, T.~Akesson {\em et.~al.}, {\it
  {Double parton scattering in $pp$ collisions at $\sqrt{s} = 63$ GeV}},  {\em
  Z. Phys.} {\bf C34} (1987) 163.

\bibitem{UA2}
{\bf UA2} Collaboration, J.~Alitti {\em et.~al.}, {\it {A Study of multi-jet
  events at the CERN $p\bar{p}$ collider and a search for double parton
  scattering}},  {\em Phys. Lett.} {\bf B268} (1991) 145--154.

\bibitem{CDF93}
{\bf CDF} Collaboration, F.~Abe {\em et.~al.}, {\it {Study of four jet events
  and evidence for double parton interactions in $p\bar{p}$ collisions at
  $\sqrt{s} = 1.8$ TeV}},  {\em Phys. Rev.} {\bf D47} (1993) 4857--4871.

\bibitem{CDF97}
{\bf CDF} Collaboration, F.~Abe {\em et.~al.}, {\it {Double parton scattering
  in $\bar{p}p$ collisions at $\sqrt{s} = 1.8 $TeV}},  {\em Phys. Rev.} {\bf
  D56} (1997) 3811--3832.

\bibitem{E735}
{\bf E735} Collaboration, T.~Alexopoulos {\em et.~al.}, {\it {The role of
  double parton collisions in soft hadron interactions}},  {\em Phys. Lett.}
  {\bf B435} (1998) 453--457.

\bibitem{D003}
{\bf D0} Collaboration, V.~M. Abazov {\em et.~al.}, {\it {Multiple jet
  production at low transverse energies in $p\bar{p}$ collisions at $\sqrt{s} =
  1.8$ TeV}},  {\em Phys. Rev.} {\bf D67} (2003) 052001,
  [hep-ex/0207046].

\bibitem{D0_2010}
{\bf D0} Collaboration, V.~M. Abazov {\em et.~al.}, {\it {Double parton
  interactions in photon+3 jet events in $p\bar{p}$ collisions at
  $\sqrt{s}=1.96$ TeV}},  {\em Phys. Rev.} {\bf D81} (2010) 052012,
  [arXiv:0912.5104].

\bibitem{D0_2011}
{\bf D0} Collaboration, V.~M. Abazov {\em et.~al.}, {\it {Azimuthal
  decorrelations and multiple parton interactions in photon+2 jet and photon+3
  jet events in $p\bar{p}$ collisions at $\sqrt{s}=1.96$ TeV}},  {\em Submitted
  to Phys. Rev. D} (2011) [arXiv:1101.1509].

\bibitem{ZEUS}
{\bf ZEUS} Collaboration, C.~Gwenlan, {\it {Multijets in photoproduction at
  HERA}},  {\em Acta Phys. Polon.} {\bf B33} (2002) 3123--3128.

\bibitem{H1}
{\bf H1} Collaboration, F.~D. Aaron, {\it {Study of multiple interactions in
  photoproduction at HERA}},
  H1-prelim-08-036.

\bibitem{TH3}
T.~Sjostrand and M.~van Zijl, {\it {A Multiple Interaction Model for the Event
  Structure in Hadron Collisions}},  {\em Phys. Rev.} {\bf D36} (1987) 2019.

\bibitem{WH}
A.~Del~Fabbro and D.~Treleani, {\it {A double parton scattering background to
  Higgs boson production at the LHC}},  {\em Phys. Rev.} {\bf D61} (2000)
  077502, [hep-ph/9911358].

\bibitem{WH1}
A.~Del~Fabbro and D.~Treleani, {\it {Double parton scatterings in b-quark pairs
  production at the LHC}},  {\em Phys. Rev.} {\bf D66} (2002) 074012,
  [hep-ph/0207311].

\bibitem{Huss}
M.~Y. Hussein, {\it {Double parton scattering in associate Higgs boson
  production with bottom quarks at hadron colliders}},
  [arXiv:0710.0203].

\bibitem{Berger_dp}
E.~L. Berger, C.~B. Jackson, and G.~Shaughnessy, {\it {Characteristics and
  Estimates of Double Parton Scattering at the Large Hadron Collider}},  {\em
  Phys. Rev.} {\bf D81} (2010) 014014,
  [arXiv:0911.5348].

\bibitem{PYT}
T.~Sjostrand, S.~Mrenna, and P.~Z. Skands, {\it {PYTHIA 6.4 Physics and
  Manual}},  {\em JHEP} {\bf 05} (2006) 026,
  [hep-ph/0603175].

\bibitem{Sjost}
T.~Sjostrand and P.~Z. Skands, {\it {Multiple interactions and the structure of
  beam remnants}},  {\em JHEP} {\bf 03} (2004) 053,
  [hep-ph/0402078].

\bibitem{WH_plb}
{\bf D0} Collaboration, V.~M. Abazov {\em et.~al.}, {\it {Search for WH
  associated production with 5.3 fb-1 of RunII data}},  {\em Submitted to Phys.
  Lett. B} (2010) [arXiv:1012.0874].

\bibitem{blazey}
G.~C. Blazey {\em et.~al.}, {\it {Run II jet physics}},
  [hep-ex/0005012].

\bibitem{fastjet}
M.~Cacciari and G.~P. Salam, {\it {Dispelling the $N^{3}$ myth for the $k_t$
  jet-finder}},  {\em Phys. Lett.} {\bf B641} (2006) 57--61,
  [hep-ph/0512210].

\bibitem{dijet}
{\bf D0} Collaboration, V.~M. Abazov {\em et.~al.}, {\it {Measurement of the
  dijet invariant mass cross section in proton anti-proton collisions at
  $\sqrt{s}$ = 1.96 TeV}},  {\em Phys. Lett.} {\bf B693} (2010) 531--538,
  [arXiv:1002.4594].

\bibitem{WZ_meas}
{\bf D0} Collaboration, V.~M. Abazov {\em et.~al.}, {\it {Measurement of the
  cross section for W and Z production to electron final state with the D0
  detector at $\sqrt{s}$ = 1.96 TeV}},
  \href{http://xxx.lanl.gov/abs/\url{http://www-d0.fnal.gov/Run2Physics/WWW/re%
sults/prelim/EW/E06/E06.pdf}} {http://www-d0.fnal.gov/Run2Physics/WWW/results/prelim/EW/E06/E06.pdf}.

\bibitem{WZ_meas1}
{\bf CDF} Collaboration, D.~Acosta {\em et.~al.}, {\it {First measurements of
  inclusive $W$ and $Z$ cross sections from Run II of the Tevatron collider}},
  {\em Phys. Rev. Lett.} {\bf 94} (2005) 091803,
  [hep-ex/0406078].

\bibitem{HW_nnlo}
T.~Hahn, S.~Heinemeyer, F.~Maltoni, G.~Weiglein, and S.~Willenbrock, {\it {SM
  and MSSM Higgs boson production cross sections at the Tevatron and the LHC}},
  [hep-ph/0607308].

\bibitem{Mikka}
{\bf D0} Collaboration, V.~M. Abazov {\em et.~al.}, {\it {Measurement of the
  inclusive jet cross-section in $p \bar{p}$ collisions at $\sqrt{s} =1.96$
  TeV}},  {\em Phys. Rev. Lett.} {\bf 101} (2008) 062001,
  [arXiv:0802.2400].

\bibitem{trf}
{\bf D0} Collaboration, V.~M. Abazov {\em et.~al.}, {\it {$b$-Jet
  Identification in the D0 Experiment}},  {\em Nucl. Instrum. Meth.} {\bf A620}
  (2010) 490--517, [arXiv:1002.4224].

\bibitem{CDF_Wj}
{\bf CDF} Collaboration, T.~Aaltonen {\em et.~al.}, {\it {Measurement of the
  cross section for $W$-boson production in association with jets in $p\bar{p}$
  collisions at $\sqrt{s}=1.96$ TeV}},  {\em Phys. Rev.} {\bf D77} (2008)
  011108(R), [arXiv:0711.4044].

\bibitem{Landsh}
P.~V. Landshoff and J.~C. Polkinghorne, {\it {Calorimeter triggers for hard
  collisions}},  {\em Phys. Rev.} {\bf D18} (1978) 3344.

\bibitem{Goebel}
C.~Goebel, F.~Halzen, and D.~M. Scott, {\it {Double Drell-Yan annihilations in
  hadron collisions: novel tests of the constituent picture}},  {\em Phys.
  Rev.} {\bf D22} (1980) 2789.

\bibitem{TH1}
F.~Takagi, {\it {Multiple production of quark jets off nuclei}},  {\em Phys.
  Rev. Lett.} {\bf 43} (1979) 1296.

\bibitem{TH11}
N.~Paver and D.~Treleani, {\it {Multi - quark scattering and large P(T) jet
  production in hadronic collisions}},  {\em Nuovo Cim.} {\bf A70} (1982) 215.

\bibitem{TH2}
B.~Humpert, {\it {Are there multi - quark interactions?}},  {\em Phys. Lett.}
  {\bf B131} (1983) 461.

\bibitem{TH21}
B.~Humpert and R.~Odorico, {\it {Multiparton scattering and QCD radiation as
  sources of four jet events}},  {\em Phys. Lett.} {\bf B154} (1985) 211.

\bibitem{Mang}
M.~L. Mangano, {\it {Four Jet Production at the Tevatron Collider}},  {\em Z.
  Phys.} {\bf C42} (1989) 331.

\bibitem{RunI_34jets}
{\bf D0} Collaboration, S.~Abachi {\em et.~al.}, {\it {Studies of Topological
  Distributions of the Three- and Four-Jet Events in $\bar{p}p$ Collisions at
  $\sqrt{s}=1800$ GeV with the D0 Detector}},  {\em Phys. Rev.} {\bf D53}
  (1996) 6000--6016, [hep-ex/9509005].

\bibitem{JN}
C.~Peterson, T.~Rognvaldsson, and L.~Lonnblad, {\it {JETNET 3.0: A Versatile
  artificial neural network package}},  {\em Comput. Phys. Commun.} {\bf 81}
  (1994) 185--220.

\end{thebibliography}

\end{document}